\documentstyle[12pt,epsfig]{article}
\makeatletter
\def\setb@se#1{\baselineskip=#1 \normalbaselineskip=#1}
\setb@se{12pt}
\long\def\title#1{\vspace*{11.5pc}{\pretolerance=10000\raggedright
  \setb@se{12pt}\bf #1\par}\nobreak\ignorespaces}
\long\def\author#1{\vspace{4pc}\begin{list}{\hfill}%
{\topsep=0pt\parskip=0pt\parsep=0pt\partopsep=0pt\listparindent=0pt%
\itemsep=0pt\rightmargin=0pt\labelsep=0pt\labelwidth=5pc\leftmargin=5pc}%
\item\normalsize{#1}\end{list}\vspace{14pt}}
\long\def\affil#1{\begin{list}{\hfill}%
{\topsep=0pt\parskip=0pt\parsep=0pt\partopsep=0pt\listparindent=0pt%
\itemsep=0pt\rightmargin=0pt\labelsep=0pt\labelwidth=5pc\leftmargin=5pc}%
  \item\normalsize{\rm #1}\end{list}\vspace{7pt}}
\long\def\beginabstract{\vspace{21pt plus 7pt minus 7pt}\begin{list}{\hfill}%
{\topsep=0pt\parskip=0pt\parsep=0pt\partopsep=0pt\listparindent=0pt%
\itemsep=0pt\rightmargin=0pt\labelsep=0pt\labelwidth=5pc\leftmargin=5pc}%
\item\normalsize{\bf Abstract. }}
\long\def\endabstract{\end{list}\vspace{28pt plus14pt minus 14pt}%
\normalsize\noindent}
\def\@sect#1#2#3#4#5#6[#7]#8{\ifnum #2>\c@secnumdepth
  \def\@svsec{}\else
  \refstepcounter{#1}\edef\@svsec{\csname the#1\endcsname.\hskip 0.5em}\fi
  \@tempskipa #5\relax
   \ifdim \@tempskipa>\z@
     \begingroup #6\relax
       \@hangfrom{\hskip #3\relax\@svsec}{\interlinepenalty \@M #8\par}%
     \endgroup
    \csname #1mark\endcsname{#7}\addcontentsline
      {toc}{#1}{\ifnum #2>\c@secnumdepth \else
                   \protect\numberline{\csname the#1\endcsname}\fi
                 #7}\else
     \def\@svsechd{#6\hskip #3\@svsec #8\csname #1mark\endcsname
                   {#7}\addcontentsline
                        {toc}{#1}{\ifnum #2>\c@secnumdepth \else
                          \protect\numberline{\csname the#1\endcsname}\fi
                    #7}}\fi
  \@xsect{#5}}

\def\section{\@startsection {section}{1}{\z@}{-28pt plus -14pt minus
-14pt}{2.3ex plus .2ex}{\normalsize\bf}}
\def\subsection{\@startsection{subsection}{2}{\z@}{-14pt plus -8pt
minus -4pt}{1.5ex plus .2ex}{\normalsize\bf}}
\def\subsubsection{\@startsection{subsubsection}{3}{\z@}{-14pt plus -8pt minus -4pt}{-1.5ex plus -.2ex}{\normalsize\bf}}
\def\paragraph{\@startsection
     {paragraph}{4}{\z@}{3.25ex plus 1ex minus .2ex}{-1em}{\normalsize\bf}}
\def\subparagraph{\@startsection
     {subparagraph}{4}{\parindent}{3.25ex plus 1ex minus
     .2ex}{-1em}{\normalsize\bf}}

\def\caption{\refstepcounter\@captype \@dblarg{\@caption\@captype}}
\long\def\@caption#1[#2]#3{\par\addcontentsline{\csname
 ext@#1\endcsname}{#1}{\protect\numberline{\csname
 the#1\endcsname}{\ignorespaces #2}}\begingroup
   \@parboxrestore
   \hspace*{28pt}
   \parbox{394pt}{\@makecaption{{\bf\csname fnum@#1\endcsname}}{\ignorespaces #3}}\par
\vspace{7pt}\endgroup}

\long\def\@makecaption#1#2{
   \vskip 14pt
   \setbox\@tempboxa\hbox{\small#1{\bf.} \small#2}
   \ifdim \wd\@tempboxa >\hsize   % IF longer than one line:
       \small#1{\bf.} \small#2\par            %   THEN set as ordinary paragraph.
     \else                        %   ELSE  set flush left.
       \hbox to\hsize{\box\@tempboxa\hfil}
   \fi}

\newcommand{\boldarrayrulewidth}{1pt} % Width of bold rule in tabular environment.

\def\bhline{\noalign{\ifnum0=`}\fi\hrule \@height  \boldarrayrulewidth \futurelet
   \@tempa\@xhline}

\def\@xhline{\ifx\@tempa\hline\vskip \doublerulesep\fi
      \ifnum0=`{\fi}}

\def\thebibliography#1{\section*{REFERENCES\@mkboth
  {REFERENCES}{REFERENCES}}\list
  {\hfil[\arabic{enumi}]}{\itemsep=0pt\labelsep=7pt\itemindent=-14pt
    \settowidth\labelwidth{[#1]}
    \leftmargin\labelwidth
    \advance\leftmargin\labelsep
    \advance\leftmargin -\itemindent
    \usecounter{enumi}}\setb@se{12pt}\small
    \def\newblock{\hskip .11em plus .33em minus .07em}
    \sloppy\clubpenalty4000\widowpenalty4000
    \sfcode`\.=1000\relax}
\def\references{\section*{REFERENCES\@mkboth
{REFERENCES}{REFERENCES}}\list{}{\itemsep=0pt\labelsep=0pt\itemindent=-28pt
\labelwidth=0pt\leftmargin=28pt}\setb@se{12pt}\small
\def\newblock{\hskip .11em plus .33em minus .07em}
\sloppy\clubpenalty4000\widowpenalty4000
\sfcode`\.=1000\relax}
\renewcommand{\d}{{\rm d}}
\newcommand{\e}{{\rm e}}

\newcommand{\APJ}{{\em Astrophys. J.} }

\newcommand{\PR}{{\em Phys. Rev.} }

\makeatother

\begin{document}
\title{LISTENING FOR RINGING BLACK HOLES}
\author{Jolien D. E. Creighton}
\affil{%
  Theoretical Astrophysics, California Institute of Technology,\\
  Pasadena CA 91125 USA
}
\abstract{%
  The gravitational radiation produced by binary black holes during
  their inspiral, merger, and ringdown phases is a promising candidate
  for detection by the first or second generation of kilometer-scale
  interferometric gravitational wave
  antennas.  Waveforms for the last phase, the quasinormal ringing,
  are well understood.  I discuss the feasibility of detection of
  the quasinormal ringing of a black hole based on an analysis of
  the Caltech 40-meter interferometer data.
}

\section{INTRODUCTION}
\label{s:intro}

Broad band interferometric gravitational wave observatories such as LIGO and
VIRGO will soon be in operation.  In order to analyze the data produced by
these instruments, one needs to know what kind of gravitational wave signals
may be present.  Since these signals are expected to be very weak, the
signal detection process requires that one know the form of the signals very
accurately in order to distinguish them from the noise in the instrument.
Additionally, one needs to understand the characteristics of the noise itself.
I describe a potentially interesting source of gravitational
radiation---the ``ringdown'' of a perturbed black hole---and discuss some of
the difficulties in searching for this radiation.

In section~\ref{s:ringdown}, I describe what a black hole ringdown is and why
it is potentially observable.  In section~\ref{s:detect}, I describe a
detection strategy that may be used to search for such a waveform in an
interferometer output.  My discussion of the statistics of the detection
strategy involves assumptions about the noise present in the interferometer.
The application of the detection strategy to real interferometer data obtained
from the Caltech 40-meter prototype is described in
section~\ref{s:observations}; I find that the observed receiver statistic
has a different distribution than the expected distribution found in
section~\ref{s:detect}.  I give a brief discussion of the implications of
this result in the final section.

\section{BLACK HOLE RINGDOWN}
\label{s:ringdown}

\subsection{Phases of black hole binary evolution}

Black hole binaries will provide an important source of gravitational radiation
for detection by LIGO and VIRGO.  The detectable radiation produced by these
systems occurs in three phases.  The first phase is the late stages of the
binary inspiral where the binary system loses energy due to gravitational
radiation reaction, and the companions spiral towards each other in a
quasi-stationary process.  Eventually, the orbit of the companions becomes
unstable, and the two black holes will plunge into each other; this merger
is the second phase.  The two black holes combine to form
a single distorted event horizon, and the distortion is dissipated into
gravitational radiation.  In time, the distortion will be well described by
linear perturbations of the final equilibrium black hole spacetime; such
perturbations are called the quasinormal modes of the final black hole.
These quasinormal modes are decaying modes: the radiation emitted from the
modes is called black hole ringdown---the third phase of gravitational
wave emission from the binary coalescence.

The relative importance of these three phases has been discussed in detail
by Hughes and Flanagan~\cite{hf:1997}.  Intermediate mass binaries
with system masses of hundreds of solar masses will produce most of their
observable signal in the merger and ringdown phases.  Merger waveforms
are not yet known, so the best way to detect such systems might
be to look for the ringdown signature.  Lower mass binaries will be best
detected by the radiation produced by the inspiral phase; however, the very
late stages of the inspiral waveforms are not yet known reliably because of
the breakdown of the post-Newtonian analysis when the system enters a regime
of strong gravity and high velocity motion.  Thus, the information gained
from the ringdown phase may be useful for binaries where the total mass is as
low as fifty solar masses.  Such binary systems could be the first sources
detected by LIGO and VIRGO: the large system mass means that these sources will
be relatively bright, which compensates for the lower event rate of black hole
binary mergers compared to neutron star binary mergers in a given volume of
space~\cite{lpp:1997}.  In addition, the search strategy I adopt here can
be used in analyzing LISA data in search of gravitational waves arising from
the ringdown of perturbed supermassive black holes.

\subsection{The ringdown phase}

The quasinormal modes of a Kerr black hole are eigenfunctions of the
Teukolsky equation---which describes linear perturbations of the curvature
of the Kerr spacetime---with boundary conditions corresponding to purely
ingoing radiation at the event horizon and purely outgoing radiation at
large distances from the black hole.  The perturbation of the curvature
of Kerr spacetime can be described by the components $\Psi_0$ and~$\Psi_4$ of
the Weyl tensor; of particular interest is $\Psi_4$ since it
describes outgoing waves in the radiative zone.  The $\Psi_4$ component is a
function of the radius~$r$, inclination~$\mu=\cos\theta$, and azimuth~$\phi$
of the observer; it also depends on the
mass~$M$ and specific angular momentum~$a$ of the perturbed Kerr black hole.
I often refer to the dimensionless angular momentum parameter~$\hat{a}=a/M$,
which must be between zero (Schwarszchild limit) and unity (extreme Kerr
limit).  Teukolsky~\cite{t:1973} was able to separate the Einstein equations,
linearized about Kerr spacetime, to obtain solutions of the form
$\Psi_4=(r-i\mu a)^{-4}\e^{-i\omega t}\,
  {}_{-2}R_{\ell m}(r)\,{}_{-2}S_{\ell m}(\mu)\,\e^{im\phi}$
where ${}_{-2}R_{\ell m}(r)$ is a solution to a radial (ordinary) differential
equation, and ${}_{-2}S_{\ell m}(\mu)$ is a spin-weighted spheroidal wave
function.  The perturbation has the spheroidal eigenvalues $\ell$ and~$m$ and
a complex frequency~$\omega$.

When the correct boundary conditions are imposed, the quasinormal modes of the
black hole are found.  These modes have a spectrum of complex
eigenfrequencies~$\omega_n$.  I shall only consider the fundamental ($n=0$)
mode because the harmonics of this mode have shorter lifetimes; for the same
reason, I shall only consider the quadrupole ($\ell=2$ and~$m=2$) mode because
this mode is the longest lived.  I assume that the observer is at a large
distance from the black hole; in this case one can approximate the
perturbation as
\begin{equation}
  \Psi_4 \approx
    \frac{A}{r}\,
    \e^{-i\omega t_{\mathrm{\scriptscriptstyle ret}}}\,
    {}_{-2}S_{22}(\mu)\,
    \e^{2i\phi}.
  \label{e:pert}
\end{equation}
Here, $t_{\mathrm{\scriptstyle ret}}=t-r^\star$ is the retarded time, where
$r^\star$ is the ``tortoise'' radial coordinate.  The parameter~$A$ represents
the amplitude of the perturbation.  The eigenfrequency spectrum can be
computed for any mode $(\ell,m,n)$: it depends on the mass and angular
momentum of the black hole.  Because the eigenfrequency is complex, the
perturbation corresponds to an exponentially damped sinusoid with a central
frequency~$f=2\pi{\mathrm{Re}}\,\omega$ and a
quality~$Q=-\frac{1}{2}{\mathrm{Re}}\,\omega/{\mathrm{Im}}\,\omega$.
For the fundamental quadrupolar mode, the
eigenfrequency is well approximated by the analytic form found by
Echeverria~\cite{e:1989}:
\begin{eqnarray}
  f &\approx& 32\,{\mathrm{kHz}}
    \times [1 - 0.63(1 - \hat{a})^{3/10}]\biggl(\frac{M_\odot}{M}\biggr) \\
  Q &\approx& 2(1 - \hat{a})^{-9/20}.
\end{eqnarray}
For a Kerr black hole with $M=50\,M_\odot$
and~$\hat{a}=0.98$, $f\approx515\,{\mathrm{Hz}}$ and~$Q\approx12$.

\subsection{Gravitational radiation}

From the curvature perturbation $\Psi_4$, one can extract useful physical
quantities.  The first quantity is the gravitational strain of the radiation.
The ``$+$'' and ``$\times$'' polarizations of the strain induced by the
gravitational radiation are found from
\begin{equation}
  h_+ - ih_\times = - \frac{2\Psi_4}{|\omega|^2}.
\end{equation}
The quantity $h_+=h_{\hat{\theta}\hat{\theta}}=h_{\hat{\phi}\hat{\phi}}$ is
the metric perturbation representing the linear polarization state along the
unit vectors ${\mathbf{e}}_{\hat{\theta}}$ and~${\mathbf{e}}_{\hat{\phi}}$,
and the quantity $h_\times=h_{\hat{\theta}\hat{\phi}}$ is the metric
perturbation representing the linear polarization state
along~${\mathbf{e}}_{\hat{\theta}}\pm{\mathbf{e}}_{\hat{\phi}}$.  The second
useful quantity that can be obtained from~$\Psi_4$ is the power radiated
(per unit solid angle) towards the observer:
\begin{equation}
  \frac{\d^2E}{\d t\,\d\Omega} =
    \lim_{r\to\infty} \frac{r^2|\Psi_4|^2}{4\pi|\omega|^2}.
  \label{e:power}
\end{equation}
Given equation~(\ref{e:pert}) for the gravitational perturbation in the far
field zone, we can integrate equation~(\ref{e:power}) over the entire sphere
and the interval $0\le t_{\mathrm{\scriptstyle ret}}<\infty$ to obtain an
expression for the total energy radiated as a function of the perturbation
amplitude.  It will be useful to characterize the amplitude of the
perturbation in terms of the fractional mass loss in the
perturbation~$\epsilon=E/M$.

I now translate the metric perturbation of a quasinormal mode into an
experimentally more useful quantity: the strain produced in an interferometric
gravitational wave antenna.  This strain is given by
$h(t_{\mathrm{\scriptstyle ret}})=
  F_+h_+(t_{\mathrm{\scriptstyle ret}})+
  F_\times h_\times(t_{\mathrm{\scriptstyle ret}})$
where $F_+$ and~$F_\times$ are the antenna response patterns of the
interferometer.  These response patterns depend on the the altitude and
azimuth of the source of the radiation as well as on the angle of polarization
of the radiation.  The exact form of theses patterns can be found in
reference~\cite{t:1987}.  For a source at a given distance, the ``typical''
waveform can be obtained by rms averaging over these angles as well as over the
inclination of the source and azimuth of the perturbation.
When this is done, one finds
\begin{eqnarray}
  h_{\mathrm{\scriptstyle ave}}(t_{\mathrm{\scriptstyle ret}}) &\approx&
    6.825\times10^{-21} \eta(\hat{a})
    \biggl(\frac{\mathrm{Mpc}}{r}\biggr)
    \biggl(\frac{M}{M_\odot}\biggr)
    \biggl(\frac{\epsilon}{0.01}\biggr)^{1/2} \nonumber\\
    &&\quad\times\: \e^{-\pi ft_{\mathrm{\scriptscriptstyle ret}}/Q}\,
    \cos(2\pi ft_{\mathrm{\scriptstyle ret}} + \psi_0).
\end{eqnarray}
where $\psi_0$ is the initial phase of the waveform and~$\eta(\hat{a})$ is
an efficiency factor which monotonically decreases from unity at $\hat{a}=0$
to zero at~$\hat{a}=1$ (note that for large values of~$\hat{a}$, $Q$ is also
large and the duration of the ringdown is long).  The efficiency remains
relatively large until the black hole has nearly extreme spin; even for a
value of~$\hat{a}=0.98$, the efficiency is reasonably high
with~$\eta(0.98)\simeq0.29$.  A Galactic black hole ringdown with
mass $M=50M_\odot$, spin $\hat{a}=0.98$, fractional
mass loss of~$\epsilon=1\%$ and distance of~$10\,{\mathrm{kpc}}$ should be
easily detected by the Caltech 40-meter prototype interferometer.

\section{SIGNAL DETECTION}
\label{s:detect}

In order to detect the ringdown from a black hole, one must pass the
interferometer output through a receiver that will perform a test
of the hypotheses ``there is a signal present in the data'' and ``there
is no signal present in the data.''  In order for the receiver to conclude
that there is a signal present, it constructs some statistic and compares
the statistic to a pre-assigned threshold.  The problem of reception,
then, is two-fold: one must find an optimal statistic, and one must select
some threshold.  My exploration of these problems (below) follows the method
presented in reference~\cite{f:1992}.

In designing the optimal statistic, it is customary to assume that the noise
in the detector is stationary and Gaussian.  These assumptions simplify the
statistical analysis.  I shall also make these assumptions about the noise
in this section; however, they are known to be poor assumptions in the case
of the Caltech 40-meter prototype interferometer.  The effect of the
non-stationary and non-Gaussian noise components will be evident in the
observations presented in section~\ref{s:observations}.

\subsection{The optimal filter}

Suppose that the detector output, $h$, contains either noise alone, $h=n$,
or both a signal and noise, $h=s+n$.  The \emph{optimal receiver} (i.e., the
optimal data analysis process) is one
that returns the quantity $P(s\mid h)$: the probability of a signal being
present given the output.  Using Bayes' law, this probability can be expressed
in terms of the \emph{a posteriori} probabilities of obtaining the output
given that a signal is or is not present, $P(h\mid s)$ and $P(h\mid\neg s)$,
and the \emph{a priori} probability of a signal being present $P(s)$ and
its converse $P(\neg s)=1-P(s)$.  One finds
\begin{equation}
  \frac{P(s\mid h)}{P(\neg s\mid h)} = 
    \frac{P(h\mid s)}{P(h\mid\neg s)}
    \frac{P(s)}{P(\neg s)} =
    \Lambda\,\frac{P(s)}{P(\neg s)}.
\end{equation}
In general, there is no universal
way of evaluating the \emph{a priori} probabilities $P(s)$ and~$P(\neg s)$, so
one often adopts the \emph{maximum likelihood receiver} which returns
the likelihood ratio $\Lambda=P(h\mid s)/P(h\mid\neg s)$.  Notice that as
$\Lambda$ grows larger, the probability of a signal increases, so the
maximum likelihood receiver can be used to test the hypotheses as follows:
If $\Lambda$ is greater than some threshold $\Lambda_\ast$ then one decides
that there is a signal present; otherwise, one decides that there is no signal
present.  Lacking any \emph{a priori} information
about whether there is a signal present, the threshold should be chosen by
setting a desired probability for a false alarm and/or a false dismissal;
these probabilities are computed in subsection~\ref{ss:properties} below.

Consider the case in which one is searching for a ringdown waveform of some
fixed frequency and quality, so it has an exactly known form.  Assume that the
noise samples are drawn from a stationary Gaussian distribution with
correlations amongst the noise events (coloured noise).  The noise correlations
can be expressed in terms of the one-sided noise power spectrum,
$\frac{1}{2}\,S_h(|f|)\delta(f-f')=
  \langle\tilde{n}(f)\tilde{n}^\ast(f')\rangle$, where $\tilde{n}(f)$ is the
Fourier transform of the noise~$n(t)$, and $\ast$ denotes complex conjugation.
Because the noise is Gaussian, the probability of obtaining an instance of
noise, $n(t)$, is $p(n)\propto\exp[-\frac{1}{2}(n\mid n)]$, where the
inner product~$(\cdot\mid\cdot)$ is defined by
\begin{equation}
  (a\mid b) = \int_{-\infty}^\infty \d f\,
    \frac{\tilde{a}^\ast(f)\tilde{b}(f)+\tilde{a}(f)\tilde{b}^\ast(f)}
         {S_h(|f|)}.
\end{equation}
If, however, a signal is present, then $h(t)=As(t)+n(t)$ where I have assumed
that the signal~$s(t)$ is normalized to some fiducial distance so that the
amplitude~$A$ represents the inverse distance of the source relative to this
fiducial distance.  The likelihood ratio is the ratio of the probabilities
$P(h\mid As)=p(h-As)$ and $P(h\mid\neg s)=p(h)$:
\begin{equation}
  \Lambda = \e^{Ax-A^2\sigma^2/2}
\end{equation}
where $x=(h\mid s)$ and $\sigma^2=(s\mid s)$.  Notice that the likelihood
ratio is a monotonically increasing function of~$x$ and that the output~$h$
appears only in the construction of~$x$; therefore, one can set a threshold
on the value of $x$ obtained rather than on the likelihood ratio.  In fact,
it will be useful to consider the signal-to-noise ratio, which I define
as $\rho=|x|/\sigma$ (the absolute value is taken because it is not known
whether the signal has a positive or a negative amplitude).

\subsection{Properties of the optimal filter}
\label{ss:properties}

It is straightforward to compute false alarm and false dismissal probabilities
for any choice of threshold and signal amplitude.  Suppose that a
threshold~$\rho_\ast$ for the signal-to-noise ratio is chosen.  Then the
false alarm probability is the probability that $\rho\ge\rho_\ast$ when no
signal is present:
\begin{equation}
  P(\mbox{false alarm}) = P(\rho\ge\rho_\ast\mid\neg s) =
   {\mathrm{erfc}}(\rho_\ast/\surd 2)
  \label{e:falsealarmA}
\end{equation}
where the complementary error function is defined
by~${\mathrm{erfc}}(x)=(2/\surd\pi)\int_x^\infty\e^{-t^2}\d t$.  When a
signal is present with amplitude~$A$, then the converse of the false dismissal
probability is the probability of a true detection:
\begin{eqnarray}
  P(\mbox{true detection}) &=& P(\rho\ge\rho_\ast\mid As) \nonumber\\
    &=& {\textstyle\frac{1}{2}}\,
    {\mathrm{erfc}}[(\rho_\ast-A\sigma)/\surd2] +
    {\textstyle\frac{1}{2}}\,
    {\mathrm{erfc}}[(\rho_\ast+A\sigma)/\surd2].
\end{eqnarray}
Notice that the signal shifts the signal-to-noise probability distribution
by~$A\sigma$.  Using these equations, one can compute the threshold required
for a choice of false alarm probability or false dismissal probability.

In the above discussion, I have made the implicit assumption that we know
the arrival time of the signal.  Since this will not be known in general,
it is necessary to obtain the signal-to-noise ratio maximized over all possible
signal arrival times.  This can be done simply by replacing the single value
of~$x$ used above by the time series obtained by the correlation
\begin{equation}
  x(t) = \int_{-\infty}^\infty \d f\,\e^{-2\pi ift}\,
    \frac{\tilde{a}^\ast(f)\tilde{b}(f)+\tilde{a}(f)\tilde{b}^\ast(f)}
         {S_h(|f|)}.
\end{equation}
The signal-to-noise ratio is then constructed by finding the maximum absolute
value of the time series~$x(t)$: $\rho=\sigma^{-1}\max_t|x(t)|$.

Unfortunately, the determination of the false alarm and true detection
probabilities are
greatly complicated because of correlations that are present in the time
series~$x(t)$.  An overestimate of the false alarm probability can be made
by assuming that $x(t)$ and~$x(t+\Delta)$ are independent where $\Delta^{-1}$
is the sampling rate.  In an observation time~$T$ consisting of
$N=T\Delta^{-1}$~samples, the probability of a false alarm is approximately
\begin{equation}
  P(\mbox{false alarm}) \approx N\,
    {\mathrm{erfc}}(\rho_\ast/\surd2) \qquad (\rho_\ast\gg1).
  \label{e:falsealarmB}
\end{equation}
for sufficiently short observation times (so that $P(\mbox{false alarm})\ll1$).
In order to
calculate the actual false alarm rate, it is necessary to use a Monte Carlo
analysis in which a large number of noise samples is simulated, and the
fractional number of samples in which the signal-to-noise ratio exceeds a
threshold is calculated.

Since the different possible waveforms depend on the mass and spin of the
black hole (or equivalently on the central frequency and the quality of the
damped sinusoid), and since these parameters are continuous, it is necessary
to discretize the waveforms to form a ``mesh'' that covers the parameter
space sufficiently finely.  By ``sufficiently finely,'' I mean that the
degradation in the signal-to-noise ratio due to having a filter with slightly
incorrect parameters should be small.  The number of templates that will be
needed to search for all ringdown waveforms of interest with very little loss
of signal-to-noise ratio will be a few thousand~\cite{hf:1997} for the Caltech
40-meter prototype: comparable to the number needed for the binary inspiral
searches.  For simplicity, I assume
hereafter that all parameters of the signal, apart from its time of arrival,
are known.

\section{OBSERVATIONS}
\label{s:observations}

Having reviewed a possible detection strategy, and derived the expected
distribution of the detection statistic in the presence of stationary Gaussian
noise, I now examine the results of applying the detection strategy to real
interferometer data.  In November of 1994, the Caltech 40-meter prototype
interferometer was used to collect approximately 46 hours of data.
In my analysis of this data, I implemented the detection strategy discussed
in the previous section (for a single filter only) using routines which are
provided in the GRASP data analysis software package~\cite{a:1997}.

The single filter I used corresponded to the fundamental quadrupole quasinormal
mode of a Kerr black hole with a mass~$M=50M_\odot$ and
a spin $\hat{a}=98\%$ of the extreme spin.  This mode is a damped sinusoid
with a central frequency of $f\simeq510\,{\mathrm{Hz}}$ and a quality
of~$Q\simeq12$.  The central frequency of the
filter is within frequency band of the instrument: between approximately
300 and $3000\,{\mathrm{Hz}}$.
The filter was cutoff when the waveform was attenuated
by $30\,{\mathrm{dB}}$ in amplitude; the filter was about $30\,{\mathrm{ms}}$
in duration.  The sampling rate of the interferometer
was $\Delta^{-1}\simeq9.868\,{\mathrm{kHz}}$.  The data were analyzed in
segments of $2^{16}$~points, of which the first and last $2^{12}$~points of the
correlation were discarded in order to remove the effects of wrap-around from
the numerical correlation algorithm.  Thus, each data segment corresponded to
approximately $5.8\,{\mathrm{s}}$ of actual data or 57\thinspace344~points.
Only $21\,{\mathrm{h}}$ (about 13\thinspace000 segments) of data were used:
these were the segments in which the instrument was in lock and
``well behaved'' in the sense that there were no outlier data points of more
than five times the sample standard deviation for the segment.

Ringdowns produced by black hole mergers within our Galaxy may have
sufficient brightness to be detected by the prototype interferometer: a
ringdown with the same parameters as the above filter occurring near the
Galactic centre ($10\,{\mathrm{kpc}}$) that radiates 1\%
of its mass would be seen with a signal-to-noise ratio of about 50.  However,
such events would be extremely rare, so I assume that the data obtained
are representative of interferometer noise alone.  It is important
to understand the properties of the noise in developing data analysis
techniques.  In particular, it is interesting to see how well the
assumptions that the noise is stationary and Gaussian made in the previous
section apply.  To this end, I estimate the false alarm probability
as a function of the threshold applied to the segments of
57\thinspace344~points
based on: (a) the signal-to-noise ratio output for all of the segments
analyzed, and (b) the signal-to-noise ratio output
for segments generated using simulated stationary Gaussian noise with the
prototype interferometer power spectrum.

\begin{figure}[t]
\begin{center}
\epsfig{file=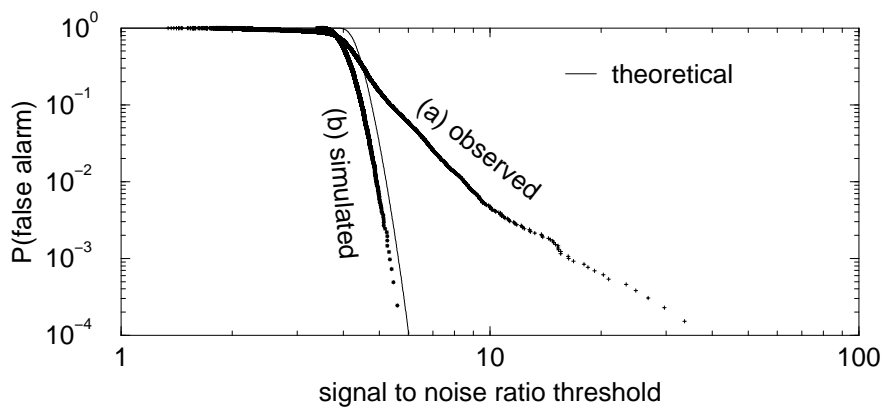,width=0.75\linewidth}
\caption{\label{f:falsealarm}The observed false alarm distributions from
  the November 1994 data run of the Caltech 40-meter prototype interferometer
  with segments lengths of 57\thinspace344~points ($\sim5.8\,{\mathrm{s}}$).
  The ``observed'' curve was obtained from the actual interferometer data,
  while the ``simulated'' curve is the expected false alarm rate for
  stationary Gaussian noise.  The solid ``theoretical'' curve corresponds to
  the theoretical false alarm distribution, equation~(\ref{e:falsealarmB}),
  under the assumption that all points in the correlation are independent.}
\end{center}
\end{figure}

These two curves are presented in figure~\ref{f:falsealarm}.  In addition
to these curves, figure~\ref{f:falsealarm} includes the false alarm
distribution (solid line) expected under the assumption that all of the
57\thinspace344~points in the correlation are independent given by
equation~(\ref{e:falsealarmB}).
Notice that the expected false alarm distribution (b) lies just to the left
of the solid line corresponding to equation~(\ref{e:falsealarmB}), as expected.
However, the observed false alarm
distribution differs greatly from the expected curve: there are an excess of
high signal-to-noise ratio events produced by the non-stationary and
non-Gaussian noise components.

\section{DISCUSSION}

Black hole ringdown waveforms will be an important potentially observable
source of gravitational radiation for the interferometric gravitational wave
detectors now being constructed.  However, the data analysis techniques to
be used to detect a black hole ringdown are just now being developed.
I have implemented the optimal filter technique to search for a single
black hole ringdown waveform---corresponding to a black hole with mass
$M=50M_\odot$ and spin $\hat{a}=98\%$ of the extreme spin---in the data
obtained from the Caltech 40-meter prototype interferometer in November~1994.
I find that there is an excess of high signal-to-noise ratio events, even
when the obviously poor data segments (those which fail an outlier test)
are excluded.

The results of the previous section indicate that the detector noise cannot
be approximated as a stationary Gaussian process for the purposes of false
alarm rate calculations.  However, there may be some method for improving
the data analysis technique that will improve the statistical properties
of the observed signal-to-noise ratios.  Of particular concern is the present
technique for estimating the power spectrum~$S_h(|f|)$ when the noise is
non-stationary.  In addition, there may be improved methods for rejecting
particularly noisy segments.  Such methods for improving the data analysis
technique are presently under investigation.

It is not surprising that the false alarm rate should be higher
than anticipated from the assumption of stationary Gaussian noise: it is known
that the prototype interferometer data contains many instrumental effects
(e.g., servo-mechanism excitations, etc.), which efficiently trigger ringdown
filters.
In order to reduce the number of false alarms, it will be necessary to develop
a set of tests to discriminate between instrumental effects and actual signals
of astronomical origin.  The simplest discriminant would be to reject signals
found in a single detector but not in other detectors in operation at that
time.  More elaborate tests will likely be needed in order to successfully
veto all the non-stationary instrumental effects; such tests are under
development.

\section*{ACKNOWLEDGEMENTS}

This work was supported by the Natural Sciences and Engineering Research
Council of Canada and by NSF grant PHY-9424337 and NASA grant
NAGW-4268/NAG5-4351.  I would like to thank the LIGO Scientific Collaboration
for providing the Caltech 40-meter prototype data for analysis, and
Kip Thorne, Albert Lazzarini, Kent Blackburn, and Barry Barish for their
comments.


\begin{thebibliography}{9}
\bibitem{hf:1997} Hughes, S. A. and Flanagan, {\'E}. {\'E}., 1997,
  \PR \emph{D} (in press).
\bibitem{lpp:1997} Lipunov, V. M., Postnov, K. A., and Prokhorov, M. E., 1997,
  \emph{New Astron.} \textbf{2} 43--52.
\bibitem{t:1973} Teukolsky, S. A., 1973, \APJ \textbf{185} 646--73.
\bibitem{e:1989} Echeverria, F., 1989, \PR \emph{D} \textbf{40} 3194--203.
\bibitem{t:1987} Thorne, K. S., 1987, in \emph{300 Years of Gravitation},
  eds S. W. Hawking and W. Israel, (Cambridge: Cambridge University Press),
  pp 330--458.
\bibitem{f:1992} Finn, L. S., 1992, \PR \emph{D} \textbf{46} 5236--48.
\bibitem{a:1997} Allen, B., 1997,
  \emph{Gravitational Radiation and Simulation Package (GRASP)},
  http://www.ligo.caltech.edu/LIGO\_web/Collaboration/manual.pdf
\end{thebibliography}
\end{document}